%% file: numuPRL2016.tex
\documentclass[aps,prl,twocolumn,showpacs,preprintnumbers,amsmath,amssymb,superscriptaddress,nofootinbib]{revtex4-1}

 \usepackage{graphicx}
 \usepackage{amsmath}
 \usepackage{amssymb}
 \usepackage{amstext}
 \usepackage{nicefrac}
 \usepackage{graphics}
 \usepackage{xspace}
 \usepackage{bm} 
 \usepackage{multirow}
 
\newcommand{\nova}{NOvA\xspace}

\newcommand{\effpot}{\ensuremath{6.05\,\mathord{\times}\,10^{20}} protons-on-target\xspace}

\newcommand{\mus}{$\mu$s\xspace}
\newcommand{\dmsq}[1]{\ensuremath{\Delta m^2_{ #1 }}\xspace}
\newcommand{\sinsq}[1]{\ensuremath{\sin^{2}\!\theta_{ #1 }}\xspace}

\newcommand{\tonetwo}{\ensuremath{\theta_{12}}\xspace}
\newcommand{\tonethree}{\ensuremath{\theta_{13}}\xspace}
\newcommand{\ttwothree}{\ensuremath{\theta_{23}}\xspace}
\newcommand{\dcp}{\ensuremath{\delta_{\rm CP}}\xspace}

\newcommand{\numu}{\ensuremath{\nu_{\mu}}\xspace}                   % nu_mu
\newcommand{\nue}{\ensuremath{\nu_{e}}\xspace}                      % nu_e
\newcommand{\nutau}{\ensuremath{\nu_{\tau}}\xspace}                 % nu_tau
\newcommand{\anue}{\ensuremath{\overline{\nu}_{e}}\xspace}          % anu_e
\newcommand{\anumu}{\ensuremath{\overline{\nu}_{\mu}}\xspace}        % anu_mu
\newcommand {\nux}[1]{\ensuremath{\nu_{ #1 }}\xspace}                 % for nu1, nu2 ...

\begin{document}
%\linenumbers


\title{Measurement of the neutrino mixing angle $\theta_{23}$ in \nova}

\input{novanumu2016.tex}  % signers

\date{\today}
\preprint{FERMILAB-PUB-17-019-ND}

\begin{abstract}
This Letter reports new results on muon neutrino disappearance from \nova, using a 14-kton detector equivalent exposure of \effpot from the NuMI beam at the Fermi National Accelerator Laboratory. The measurement probes the muon-tau symmetry hypothesis that requires maximal $\theta_{23}$ mixing ($\theta_{23} = \pi/4$). Assuming the normal mass hierarchy, we find $\dmsq{32}=(2.67\pm 0.11)\mathord{\times}10^{-3}$\,eV$^{2}$ and \sinsq{23} at the two statistically degenerate values $0.404^{+0.030}_{-0.022}$ and $0.624^{+0.022}_{-0.030}$, both at the 68\% confidence level. Our data disfavor the maximal mixing scenario with $2.6\sigma$ significance.
\end{abstract}

\maketitle

Neutrino flavor states (\nue, \numu and \nutau) are superpositions of neutrino mass eigenstates (\nux{1}, \nux{2} and \nux{3}), giving rise to the phenomenon of neutrino oscillations. The superpositions are described by the unitary matrix, $U_{\text{PMNS}}$~\cite{ref:PMNS}, that can be parameterized in terms of three mixing angles (\tonetwo, \tonethree, and \ttwothree) and a $CP$-violating phase~\dcp. For a given distance traveled, the energy at which the largest oscillation probability occurs is governed by the differences in the squared masses of the neutrinos, \dmsq{21} and \dmsq{32}. The mixing angles and mass-squared differences have been measured by multiple experiments~\cite{ref:Experiments,ref:sk,ref:MinosCombined,ref:T2KCombined,ref:numuFA}. However, considerable uncertainty remains on the value of \dcp, the sign of \dmsq{32}, and whether \ttwothree is maximal, in the upper octant, or in the lower octant ($\ttwothree=\pi/4$, $\ttwothree>\pi/4$, or $\ttwothree<\pi/4$, respectively). Should $\ttwothree=\pi/4$, the \numu and \nutau components of the $\nu_3$ mass eigenstate would be equal. Previous experimental results are compatible with $\theta_{23}=\pi/4$~\cite{ref:sk,ref:MinosCombined,ref:T2KCombined,ref:numuFA}, motivating theoretical models with an underlying muon-tau symmetry in the neutrino sector~\cite{ref:MaxMix}. More precise measurements are valuable in identifying viable theories of neutrino masses and mixing. In this Letter, we present updated measurements of \sinsq{23} and \dmsq{32} by analyzing \numu disappearance in \nova data collected between February~6, 2014 and May~2, 2016. This corresponds to an accumulated 14~kton detector equivalent exposure of \effpot, which is 2.2~times that used in our previous publication~\cite{ref:numuFA}. 

\nova is a long-baseline neutrino oscillation experiment with two functionally identical detectors~\cite{ref:nova,ref:numuFA,ref:nueFA,ref:novaElders}. The energy spectrum of the neutrinos produced by the NuMI beam~\cite{ref:NuMI} at the Fermi National Accelerator Laboratory is measured by the Near Detector (ND) located 1~km away from the NuMI target. The neutrinos are subsequently detected 810~km away in the Far Detector (FD) near Ash River, MN. The 14-kton FD is located on the surface while the 290-ton ND is 100~m underground. Both detectors are sited off the central beam axis. The FD is 14.6~mrad off-axis so that the resulting narrow neutrino-energy spectrum peaks around 2~GeV, near the first oscillation maximum. The ND is positioned to maximize the similarity between the neutrino energy spectra observed at the two detectors. The flavor composition of beam neutrinos interacting in the ND\,(FD) is estimated from simulation to be 97.5\%\,(97.8\%)~\numu, 1.8\%\,(1.6\%)~\anumu and 0.7\%\,(0.6\%)~$\nue+\anue$ between 1-3~GeV, assuming no oscillations.

Both detectors are segmented, tracking calorimeters with organic scintillator constituting 62\% of their fiducial mass. Reflective polyvinyl chloride cells~\cite{ref:PVC} of length 15.5~m~(3.9~m) in the FD~(ND) with a $3.9\times6.6$~cm$^2$ cross section are filled with liquid scintillator~\cite{ref:Scintillator}. The cells are arranged in 896~(214)~planes in the FD~(ND) and alternate between vertical and horizontal orientations to allow three-dimensional reconstruction. Muon containment is improved at the downstream end of the ND by ten layers of 10-cm-thick steel. Each layer of steel is interleaved with two planes of scintillator, one in each orientation. Light produced by charged particles is collected by a loop of wavelength-shifting optical fiber in each cell~\cite{ref:Kuraray} and measured with an avalanche photodiode~(APD)~\cite{ref:Hamamatsu}. APD signals within a 550-\mus time window centered on the 10-\mus NuMI beam spill are stored. Other time windows are also recorded for calibration and background measurements.

Precise determination of the oscillation parameters governing \numu disappearance, primarily \dmsq{32} and \sinsq{23}, requires identification of charged-current~(CC) interactions of muon neutrinos in the beam and an accurate estimate of their energy. Backgrounds from neutral current (NC), \nue-CC, and \nutau-CC interactions must be rejected along with particles originating from outside the detector, particularly cosmic rays at the FD and neutrino-induced muons at the ND. The energy of a \numu-CC interaction is estimated by summing the reconstructed energy of the muon and the hadronic recoil system. 

We use a comprehensive Monte Carlo (MC) simulation of the neutrino beam and our detectors in this analysis. Hadron production in the target is modeled using \textsc{fluka}~\cite{ref:FLUKA}, while the focusing and decay of those hadrons in the NuMI beam line is simulated using the \textsc{flugg}~\cite{ref:FLUGG} interface to \textsc{geant4}~\cite{ref:GEANT4}. Neutrino interactions are simulated using \textsc{genie}~\cite{ref:GENIE} with the modifications outlined below. Our detector simulation uses \textsc{geant4} along with custom software to model photon transport and capture in different detector elements, as well as the response of the APD and readout electronics~\cite{ref:Simulation}. 

Evidence presented by other experiments~\cite{ref:ExpMEC} suggests additional event rate and an alteration of kinematic distributions arising in neutrino scattering on nuclei. Analysis of the hadronic energy distribution in the NOvA ND data further support this conclusion. While this is an area of active theoretical development~\cite{ref:TheoMEC}, for the results presented here, our simulation has been augmented with a semi-empirical model in \textsc{genie} that posits neutrinos scatter from nucleon pairs ($np$ and $nn$) within the nucleus~\cite{ref:ModelMEC}. The model is inspired by observations of rate enhancements in electron-nucleus scattering data and their treatment via 2-particle 2-hole~($2p2h$) calculations that include meson exchange currents~(MECs)~\cite{ref:eScatRev}. Adjustments were made to the semi-empirical model to achieve a more constant cross section for $2p2h$-MEC processes above 1~GeV. These events are also reweighted as a function of three-momentum transfer and visible hadronic energy to match the ND data. The addition of $2p2h$-MEC processes increases the simulated event rate by about 10\% in both detectors, but the mean reconstructed neutrino energy and spectral shape remains largely unchanged. Additionally, as suggested by a reanalysis of bubble chamber data~\cite{ref:reBubble}, the rate of \numu-CC nonresonant single pion production in \textsc{genie} is reduced by 50\%~\cite{ref:footnoteDIS}.

Our data analysis starts with a collection of cells that have an APD signal above threshold.  These hits are then clustered in space and time~\cite{ref:slicer} to construct event candidates. Trajectories of charged particles are reconstructed using a technique based on the Kalman filter algorithm~\cite{ref:Kalman}. The resulting tracks are analyzed to identify muon candidates~\cite{ref:NickRustemTheses} by using four variables as inputs to a $k$-nearest neighbor ($k$NN) classifier~\cite{ref:kNN}: $dE/dx$ likelihood, total track length, scattering likelihood, and fraction of planes along the track consistent with having additional hadronic activity. The $k$NN classifier is applied to all tracks in an event and the track with the highest output is used to select \numu-CC candidate events. The impact of secondary particles carrying energy out of the detector is minimized by removing candidate events with hits in the outer two cells or planes, as well as events that have a short projected distance from the track ends to a detector edge. These containment requirements also significantly aid rejection of backgrounds originating outside the detector volume. The NC background is estimated from simulation to be 1.5\% of the ND sample, while the background coming from both \nue-CC and \nutau-CC is well below 1\%.

Further event selection criteria are applied to minimize the contribution from cosmic ray background in the FD. As a first step, we select events within a 12~\mus window centered on the beam spill. Additional cosmic ray rejection is achieved using a boosted decision tree~\cite{ref:BDT} that includes information on the reconstructed event topology, such as track angle with respect to the beam, fraction of hits in the track, and scattering information. A high-statistics cosmic ray data set, recorded at times when there was no beam, was used in conjunction with simulated neutrino interactions to tune the cosmic rejection criteria. Using a separate data set collected alongside the beam spills in the long 550-\mus readout window, we measure the rate of cosmic-induced background events passing our selection criteria. Overall, we reduce the cosmic-induced events occurring during the beam spills by 7 orders of magnitude, resulting in a cosmic background that is lower than the number of selected beam background events. The uncertainty on the remaining cosmic background is 9\%, due to the limited size of that sample. The efficiency in the FD simulation for selecting contained \numu-CC interactions is 62\%.

\begin{figure*}[t!]
\includegraphics[width=0.99\textwidth]{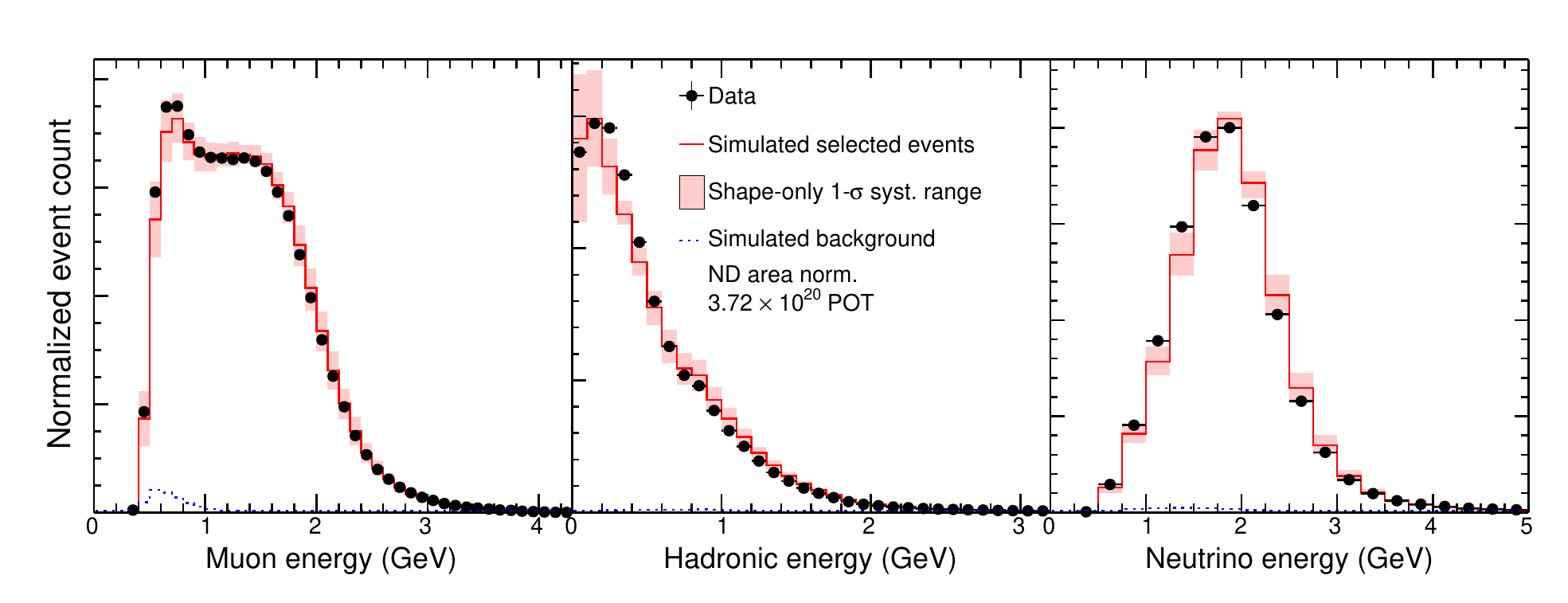}
\caption{Reconstructed muon (left), hadronic (center), and neutrino (right) energy for $1.09 \times 10^{6}$ selected \numu-CC interactions in the ND. After selection, data (black dots) and Monte Carlo (red) normalization differ by 1.1\%, which is removed from the plot by normalizing by area. The systematic error band contains only the bin-to-bin uncertainties, suppressing the 20\%$-$30\% absolute normalization uncertainties primarily due to neutrino flux and cross sections. Simulated backgrounds are shown in dotted blue.}
\label{fig:ND3plots}
\end{figure*}

Muon energy is reconstructed from the measured path length in the detector. Hadronic energy is obtained from calorimetry by first summing all the visible energy not associated with the muon. A piecewise linear fit obtained from simulation~\cite{ref:SusanThesis} is used to relate the summed visible energy to the estimated total hadronic energy. The estimated muon and hadronic energy resolution from our simulation are 3.5\% and 25\% respectively, giving an overall energy resolution for selected \numu-CC events of about 7\% for both detectors. Studies of ND data show that the energy resolution is well modeled and that any remaining differences between data and MC are accounted for by the systematic uncertainties considered in the analysis. Figure~\ref{fig:ND3plots} shows the reconstructed muon energy, hadronic energy and neutrino energy for selected \numu-CC interactions in the ND. The observed 2.6\% difference in the mean neutrino energy between data and MC is consistent with the total systematic uncertainty, as visualized by the (bin-to-bin correlated) red shaded band in Fig.~\ref{fig:ND3plots}. 

Discrepancies between data and MC in the ND energy spectrum are extrapolated to produce a predicted FD spectrum while accounting for the different flux and acceptance at each detector. In the first step of the extrapolation, we subtract the background from the ND spectrum as estimated from simulation. We convert the ND reconstructed energy spectrum into a true energy spectrum using the reconstructed-to-true migration matrix obtained from the ND simulation, and then multiply by the FD-to-ND event ratio as a function of true neutrino energy to obtain the FD true energy spectrum. The ratio also incorporates the effect of three-flavor neutrino oscillations, including matter effects, for any particular choice of the oscillation parameters. The FD true energy prediction is transformed into a reconstructed energy prediction using the simulated FD migration matrix. In the final step, the data-based cosmic and simulation-based beam-induced backgrounds (NC, \nue-CC and \nutau-CC) are added to the prediction, which is then compared to the FD data.

Our measurement of \sinsq{23} and \dmsq{32} accounts for systematic uncertainties in the energy scale, normalization, neutrino cross section and final-state interactions, neutrino flux, and backgrounds. These uncertainties can have an interdetector (relative) contribution, due to differences between the ND and the FD, and an absolute contribution that affects both detectors in the same way. The relative and absolute hadronic energy scale uncertainties are both estimated as 5\%, based on studies of the ND response to protons in data compared to simulation, and a comparison of different hadronic interaction models in \textsc{geant4}. The absolute muon energy scale uncertainty is set at 2\% based on uncertainties in the simulation of energy loss in the detector materials~\cite{ref:PDG}. The relative muon energy scale uncertainty, also 2\%, arises from uncertainties in the material composition of the ND and the FD. A relative normalization uncertainty of 5\% is dominated by the impact on the reconstruction efficiency of activity originating outside the detector. Neutrino cross section and hadronization uncertainties are taken from Ref.~\cite{ref:GENIEManual} with the following exceptions. The rescaled \numu-CC nonresonant single-pion component is assigned a 50\% uncertainty. Additionally, the $2p2h$-MEC model rate uncertainty is also taken as 50\%, motivated by remaining discrepancies between ND data and MC. The absolute neutrino flux uncertainty of approximately 20\% near the peak of the spectrum is dominated by uncertainties on hadron production~\cite{ref:NA49}. This uncertainty is strongly correlated between the two detectors and is mitigated by the extrapolation procedure. The uncertainty on the number of selected NC, \nue-CC, and \nutau-CC background events is conservatively estimated at 100\%. The simulated light output as a function of $dE/dx$ was tuned using proton and muon tracks in the ND. The difference between the tuned response and the standard parameterization~\cite{ref:Birks} was taken as a systematic uncertainty. Evaluation of different noise models in the simulation shows negligible changes to the energy scale and normalization. The main components of the analysis, including muon identification and event containment criteria, as well as muon and hadronic energy reconstruction, are nearly the same in both detectors, thereby reducing the impact of systematic effects. Table~\ref{tab:errors} summarizes the sources of uncertainty and their impact on the \sinsq{23} and \dmsq{32} measurements. The size of the impact is estimated by using the 68\%~C.L. interval from a fit to simulated data with only statistical uncertainty compared to a fit with the systematic uncertainty also included.

\begin{table*}[t]
\caption{Sources of uncertainty and their estimated average impact on the \sinsq{23} and \dmsq{32} measurements. For this table, the impact is quantified using the increase in the one-dimensional 68\%~C.L. interval, relative to the size of the interval when only statistical uncertainty is included in the fit. 
Simulated data were used and oscillated with $\dmsq{32}=2.67\mathord{\times}10^{-3}\text{~eV}^{2}$ and $\sinsq{23}=0.626$.}
\begin{tabular}{c c c c}
\hline 
\multirow{2}{*}{Source of uncertainty} & Uncertainty in &  & Uncertainty in \\
& $\sin^2\!\theta_{23} (\times 10^{-3})$ & \phantom{A} & \dmsq{32} $\left(\times 10^{-6}\text{ eV}^{2}\right)$   \\
\hline 
Absolute muon energy scale [$\pm2\%$] & +9 /  -8 & & +3 /  -10\\
Relative muon energy scale [$\pm2\%$] & +9 /  -9 & & +23 /  -14\\
Absolute hadronic energy scale [$\pm5\%$] & +5 /  -5 & & +7 /  -3\\
Relative hadronic energy scale [$\pm5\%$] & +10 /  -11 & & +29 /  -19\\
Normalization [$\pm5\%$] & +5 /  -5 & & +4 /  -8 \\
Cross sections and final-state interactions & +3 /  -3  & & +12 /  -15 \\ 
Neutrino flux & +1 /  -2 & & +4 /  -7 \\
Beam background normalization [$\pm100\%$] & +3 /  -6 & & +10 /  -16 \\ 
Scintillation model & +4 /  -3   & & +2 /  -5 \\
\dcp $(0 - 2\pi)$ & +0.2 /  -0.3 & & +10 /  -9 \\
\hline 
Total systematic uncertainty & +17 /  -19 & & +50 /  -47 \\
\hline
Statistical uncertainty & +21 /  -23 & & +93 /  -99  \\
\hline
\end{tabular}
\label{tab:errors}
\end{table*}

\begin{figure}[!htbp]
   \centering
   \includegraphics[width=0.49\textwidth]{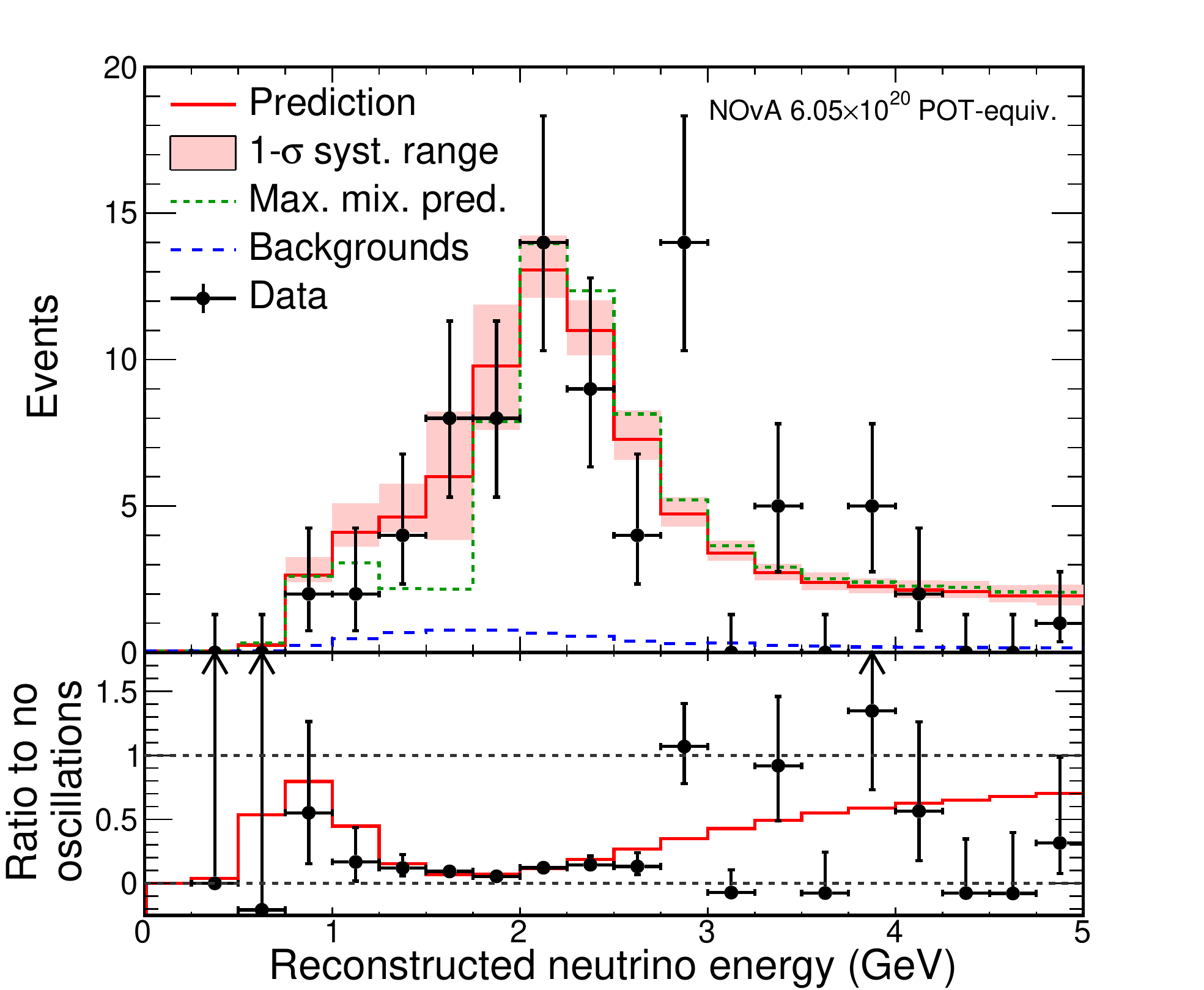} 
   \caption{Top: Comparison of the reconstructed energy spectrum of the FD data (black dots) and best-fit prediction (red). The systematic uncertainty band is shaded red. Combined beam and cosmic backgrounds are shown by the dashed blue histogram. The prediction assuming maximal mixing is shown in dashed green. Bottom: The ratio to no oscillations for data and MC after background subtraction.}
   \label{fig:FDspectrum}
\end{figure}

We performed a blind analysis where energy, muon-classifier values and the number of FD beam events were obscured until the analysis was finalized. After unblinding, we observed 78~\numu-CC candidate events in the FD with an expected background of 3.4~NC, 0.23~\nue-CC, 0.27~\nutau-CC events, and 2.7~cosmic-ray-induced events. In the absence of oscillations $473\pm30$~events are predicted. At the best-fit parameters, 82.4~events are expected. Figure~\ref{fig:FDspectrum} shows the measured energy spectrum along with the best fit prediction, with the ratio to the prediction in the absence of oscillations shown in the lower panel. The data are fit for oscillations using 19~energy bins of 0.25~GeV width between 0.25-5.0~GeV. The fit uses a log-likelihood minimization with systematic uncertainties profiled using Gaussian penalty terms. The oscillation parameters not directly measured in this analysis are also profiled over, using uncertainties taken from world averages~\cite{ref:PDG}. Our best fit is quoted at $\dcp=3\pi/2$, which is degenerate with $\dcp=\pi/2$. The disappearance probability is only mildly dependent on the value of \dcp and the effect of letting \dcp vary in the $[0, 2\pi]$ range is included in the uncertainties.

The best fit to the data gives $\dmsq{32}=(+2.67\pm 0.11)\mathord{\times}10^{-3}\text{~eV}^{2}$ and $\sinsq{23}$ at the two statistically degenerate values $0.404^{+0.030}_{-0.022}$ and $0.624^{+0.022}_{-0.030}$ both at the 68\%~C.L. in the normal hierarchy~(NH). For the inverted hierarchy, $\dmsq{32}=(-2.72\pm 0.11)\mathord{\times}10^{-3}\text{~eV}^{2}$ and $\sinsq{23} = 0.398^{+0.030}_{-0.022}$ or $0.618^{+0.022}_{-0.030}$ at 68\%~C.L. The best fit has a $\chi^2/\text{d.o.f.}=41.6/17$, which arises mainly from bins in the tail of the energy spectrum that contain little information about the three-flavor oscillations. Restricting the fit to energies below 2.5~GeV reduces the $\chi^2/\text{d.o.f.}$ to $3.2/7$ and does not significantly change the fit results. 

Maximal mixing, where $\sinsq{23}=0.5$, is disfavored by the data at $2.6\sigma$. Fixing $\sinsq{23}=0.5$ gives a best fit of $\dmsq{32}=2.48\mathord{\times}10^{-3}\text{~eV}^{2}$~(NH) with a prediction of 77.7~events. Figure~\ref{fig:FDspectrum} illustrates the difference between the energy spectrum for the maximal mixing prediction, in dashed green, and the best fit to our data, in red, for which the mixing is nonmaximal. The 1-2~GeV region is where the oscillation maximum occurs and the events in that range provide the most information about the mixing angle. Visual scanning of the events in this region along with studies of their geometric location and kinematic variables gave results consistent with expectations.

\begin{figure}[!htbp]
   \centering
   \includegraphics[width=0.49\textwidth]{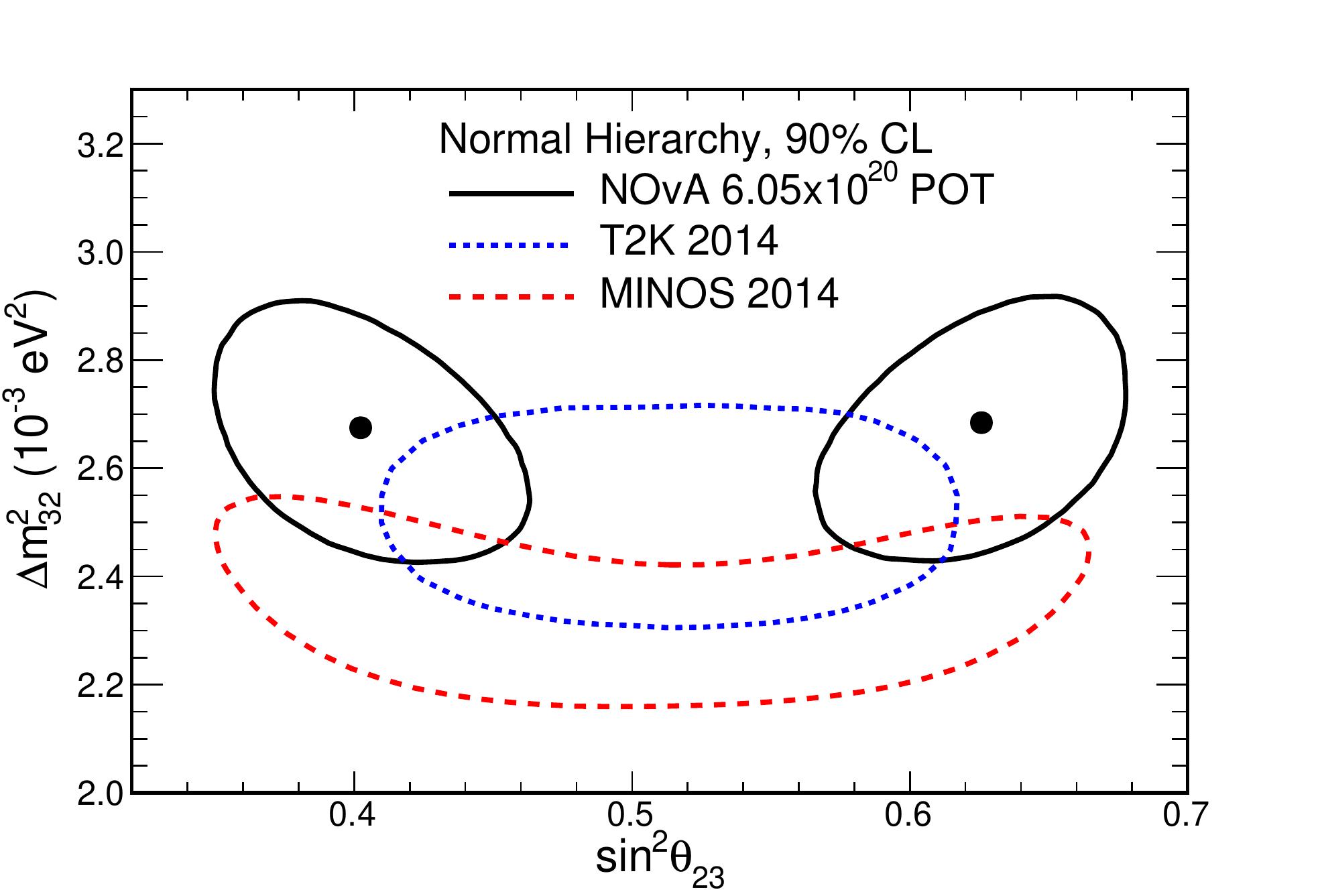} 
   \caption{Best fit (black dots) and allowed 90\%~C.L. regions (solid black curves) of \sinsq{23} and \dmsq{32} for the NH. The dashed curves show MINOS~\cite{ref:MinosCombined} and T2K~\cite{ref:T2KCombined} 90\%~C.L. contours.}
   \label{fig:contour}
\end{figure}

Figure~\ref{fig:contour} shows the allowed 90\%~C.L. regions in $\dmsq{32}$ and $\sinsq{23}$ where two islands form, one for each $\theta_{23}$ octant. The statistical significance of these contours, as well as the 68\% confidence levels for each observable, have been determined using the Feldman-Cousins unified approach~\cite{ref:FeldmanCousins}. These new results are consistent with those in our previous publication~\cite{ref:numuFA}. Contours from MINOS~\cite{ref:MinosCombined} and T2K~\cite{ref:T2KCombined} are also shown in Fig.~\ref{fig:contour} for comparison. 

In summary, using more than double the data in the previous result, \nova has observed muon neutrino disappearance and performed a high precision measurement of the oscillation parameters. Our data disfavor a value of $\theta_{23}=\pi/4$ at $2.6\sigma$ significance. 

This work was supported by the US Department of Energy; the US National Science Foundation; the Department of Science and Technology, India; the European Research Council; the MSMT CR, GA UK, Czech Republic; the RAS, RMES, and RFBR, Russia; CNPq and FAPEG, Brazil; and the State and University of Minnesota. We are grateful for the contributions of the staffs of the University of Minnesota module assembly facility and \nova FD Laboratory, Argonne National Laboratory, and Fermilab. Fermilab is operated by Fermi Research Alliance, LLC under Contract No. De-AC02-07CH11359 with the US DOE.

\textit{Note added.}- Recently, a new measurement of $\theta_{23}$ was released by the T2K Collaboration~\cite{ref:T2Kcombined}. The overlap between the contours of \nova and the updated T2K result is at a very similar level to what is shown here.

\end{document}

%% file: novanumu2016.tex
\newcommand{\ANL}{Argonne National Laboratory, Argonne, Illinois 60439,
USA}
\newcommand{\IOP}{Institute of Physics, The Czech Academy of Sciences,
182 21 Prague, Czech Republic}
\newcommand{\BHU}{Department of Physics, Institute of Science, Banaras
Hindu University, Varanasi, 221 005, India}
\newcommand{\Caltech}{California Institute of
Technology, Pasadena, California 91125, USA}
\newcommand{\Cochin}{Department of Physics, Cochin University
of Science and Technology, Kochi 682 022, India}
\newcommand{\Charles}
{Charles University, Faculty of Mathematics and Physics,
 Institute of Particle and Nuclear Physics, 116 36 Prague, Czech Republic}
\newcommand{\Cincinnati}{Department of Physics, University of Cincinnati,
Cincinnati, Ohio 45221, USA}
\newcommand{\CSU}{Department of Physics, Colorado
State University, Fort Collins, CO 80523-1875, USA}
\newcommand{\CTU}{Czech Technical University in Prague,
Brehova 7, 115 19 Prague 1, Czech Republic}
\newcommand{\Dallas}{Physics Department, University of Texas at Dallas,
800 W. Campbell Rd. Richardson, Texas 75083-0688, USA}
\newcommand{\Delhi}{Department of Physics and Astrophysics, University of
Delhi, Delhi 110007, India}
\newcommand{\JINR}{Joint Institute for Nuclear Research,
Dubna, Moscow region 141980, Russia}
\newcommand{\FNAL}{Fermi National Accelerator Laboratory, Batavia,
Illinois 60510, USA}
\newcommand{\UFG}{Instituto de F\'{i}sica, Universidade Federal de
Goi\'{a}s, Goi\^{a}nia, Goi\'{a}s, 74690-900, Brazil}
\newcommand{\Guwahati}{Department of Physics, IIT Guwahati, Guwahati, 781
039, India}
\newcommand{\Harvard}{Department of Physics, Harvard University,
Cambridge, Massachusetts 02138, USA}
\newcommand{\IHyderabad}{Department of Physics, IIT Hyderabad, Hyderabad,
502 205, India}
\newcommand{\Hyderabad}{School of Physics, University of Hyderabad,
Hyderabad, 500 046, India}
\newcommand{\Indiana}{Indiana University, Bloomington, Indiana 47405,
USA}
\newcommand{\INR}{Inst. for Nuclear Research of Russia, Academy of
Sciences 7a, 60th October Anniversary prospect, Moscow 117312, Russia}
\newcommand{\Iowa}{Department of Physics and Astronomy, Iowa State
University, Ames, Iowa 50011, USA}
\newcommand{\Irvine}{Department of Physics and Astronomy,
University of California at Irvine, Irvine, California 92697, USA}
\newcommand{\Jammu}{Department of Physics and Electronics, University of
Jammu, Jammu Tawi, 180 006, Jammu and Kashmir, India}
\newcommand{\Lebedev}{Nuclear Physics Department, Lebedev Physical
Institute, Leninsky Prospect 53, 119991 Moscow, Russia}
\newcommand{\MSU}{Department of Physics and Astronomy, Michigan State
University, East Lansing, Michigan 48824, USA}
\newcommand{\Duluth}{Department of Physics and Astronomy,
University of Minnesota Duluth, Duluth, Minnesota 55812, USA}
\newcommand{\Minnesota}{School of Physics and Astronomy, University of
Minnesota Twin Cities, Minneapolis, Minnesota 55455, USA}
\newcommand{\Panjab}{Department of Physics, Panjab University,
Chandigarh, 106 014, India}
\newcommand{\SAlabama}{Department of Physics, University of
South Alabama, Mobile, Alabama 36688, USA}
\newcommand{\Carolina}{Department of Physics and Astronomy, University of
South Carolina, Columbia, South Carolina 29208, USA}
\newcommand{\SDakota}{South Dakota School of Mines and Technology, Rapid
City, South Dakota 57701, USA}
\newcommand{\SMU}{Department of Physics, Southern Methodist University,
Dallas, Texas 75275, USA}
\newcommand{\Stanford}{Department of Physics, Stanford University,
Stanford, California 94305, USA}
\newcommand{\Sussex}{Department of Physics and Astronomy, University of
Sussex, Falmer, Brighton BN1 9QH, United Kingdom}
\newcommand{\Tennessee}{Department of Physics and Astronomy,
University of Tennessee, Knoxville, Tennessee 37996, USA}
\newcommand{\Texas}{Department of Physics, University of Texas at Austin,
Austin, Texas 78712, USA}
\newcommand{\Tufts}{Department of Physics and Astronomy, Tufts University, Medford,
Massachusetts 02155, USA}
\newcommand{\UCL}{Physics and Astronomy Dept., University College London,
Gower Street, London WC1E 6BT, United Kingdom}
\newcommand{\Virginia}{Department of Physics, University of Virginia,
Charlottesville, Virginia 22904, USA}
\newcommand{\WSU}{Department of Mathematics, Statistics, and Physics,
Wichita State Univ.,
Wichita, Kansas 67206, USA}
\newcommand{\WandM}{Department of Physics, College of William \& Mary,
Williamsburg, Virginia 23187, USA}
\newcommand{\Winona}{Department of Physics, Winona State University, P.O.
Box 5838, Winona, Minnesota 55987, USA}
\newcommand{\deceased}{Deceased.}

\affiliation{\ANL}
\affiliation{\IOP}
\affiliation{\BHU}
\affiliation{\Caltech}
\affiliation{\Charles}
\affiliation{\Cincinnati}
\affiliation{\Cochin}
\affiliation{\CSU}
\affiliation{\CTU}
\affiliation{\Delhi}
\affiliation{\FNAL}
\affiliation{\UFG}
\affiliation{\Guwahati}
\affiliation{\Harvard}
\affiliation{\Hyderabad}
\affiliation{\IHyderabad}
\affiliation{\Indiana}
\affiliation{\INR}
\affiliation{\Iowa}
\affiliation{\Irvine}
\affiliation{\Jammu}
\affiliation{\JINR}
\affiliation{\Lebedev}
\affiliation{\MSU}
\affiliation{\Duluth}
\affiliation{\Minnesota}
\affiliation{\Panjab}
\affiliation{\SAlabama}
\affiliation{\Carolina}
\affiliation{\SDakota}
\affiliation{\SMU}
\affiliation{\Stanford}
\affiliation{\Sussex}
\affiliation{\Tennessee}
\affiliation{\Texas}
\affiliation{\Tufts}
\affiliation{\UCL}
\affiliation{\Virginia}
\affiliation{\WSU}
\affiliation{\WandM}
\affiliation{\Winona}

\author{P.~Adamson}
\affiliation{\FNAL}

\author{L.~Aliaga}
\affiliation{\FNAL}

\author{D.~Ambrose}
\affiliation{\Minnesota}

\author{N.~Anfimov}
\affiliation{\JINR}

\author{A.~Antoshkin}
\affiliation{\JINR}

\author{E.~Arrieta-Diaz}
\affiliation{\SMU}

\author{K.~Augsten}
\affiliation{\CTU}

\author{A.~Aurisano}
\affiliation{\Cincinnati}

\author{C.~Backhouse}
\affiliation{\Caltech}

\author{M.~Baird}
\affiliation{\Sussex}
\affiliation{\Indiana}

% ...

\collaboration{The NOvA Collaboration}
\noaffiliation

%% file: numuPRL2016.bbl
\begin{thebibliography}{99}
\bibitem{ref:footnoteDIS} The reduction is achieved via a 35\% reduction in deep inelastic scattering events with $W<1.7$~GeV.
\end{thebibliography}
